

\hoffset=0.5in \voffset=0.2cm
\vbadness=10000
%
%

\font\bf=cmbx10 scaled 1200

\font\it=cmti10 scaled 1200

\font\tenrm=cmr10 scaled 1200
\font\sevenrm=cmr7 scaled 1200
\font\fiverm=cmr5 scaled 1200
\font\teni=cmmi10 scaled 1200
\font\seveni=cmmi7 scaled 1200
\font\fivei=cmmi5 scaled 1200
\font\tensy=cmsy10 scaled 1200
\font\sevensy=cmsy7 scaled 1200
\font\fivesy=cmsy5 scaled 1200

\font\tenbf=cmbx10 scaled 1200
\font\sevenbf=cmbx7 scaled 1200
\font\fivebf=cmbx5 scaled 1200
\font\tensl=cmsl10 scaled 1200
\font\tentt=cmtt10 scaled 1200
\font\tenit=cmti10 scaled 1200
\catcode`\@=11
\textfont0=\tenrm \scriptfont0=\sevenrm \scriptscriptfont0=\fiverm
\def\rm{\fam\z@\tenrm}
\textfont1=\teni \scriptfont1=\seveni \scriptscriptfont1=\fivei
\def\mit{\fam\@ne} \def\oldstyle{\fam\@ne\teni}
\textfont2=\tensy \scriptfont2=\sevensy \scriptscriptfont2=\fivesy
\def\cal{\fam\tw@}
\textfont3=\tenex \scriptfont3=\tenex \scriptscriptfont3=\tenex
\newfam\itfam \def\it{\fam\itfam\tenit} 
\textfont\itfam=\tenit
\newfam\slfam  
\textfont\slfam=\tensl
\newfam\bffam \def\bf{\fam\bffam\tenbf} 
\textfont\bffam=\tenbf \scriptfont\bffam=\sevenbf
\scriptscriptfont\bffam=\fivebf
\newfam\ttfam  
\textfont\ttfam=\tentt
\catcode`\@=12
\rm


\abovedisplayskip=30pt plus 4pt minus 10pt
\abovedisplayshortskip=20pt plus 4pt
\belowdisplayskip=30pt plus 4pt minus 10pt
\belowdisplayshortskip=28pt plus 4pt minus 4pt

\hfuzz=10pt \overfullrule=0pt
\vsize 8in
\hsize 6in
\baselineskip=22pt

\def\singlespace{\baselineskip=14pt}
\parindent 20pt \parskip 6pt

 \mathcode`*="002A

\def\_{\vrule height 0.8pt depth 0pt width 1em}

\newbox\grsign \setbox\grsign=\hbox{$>$} \newdimen\grdimen \grdimen=\ht\grsign
\newbox\simlessbox \newbox\simgreatbox
\setbox\simgreatbox=\hbox{\raise.5ex\hbox{$>$}\llap
     {\lower.5ex\hbox{$\sim$}}}\ht1=\grdimen\dp1=0pt
\setbox\simlessbox=\hbox{\raise.5ex\hbox{$<$}\llap
     {\lower.5ex\hbox{$\sim$}}}\ht2=\grdimen\dp2=0pt

\def\boxx{\mathop{\hbox to 0pt{$\sqcup$\hss}\hbox to
     0pt{$\sqcap$\hss}\phantom\nabla}}


\baselineskip=22pt

\singlespace


\centerline{\bf Gravitational Radiation}
\centerline{\bf and Very Long Baseline Interferometry}
\vskip .5cm

\centerline{Ted Pyne,$^{(1)}$ Carl R. Gwinn,$^{(2)}$, Mark Birkinshaw,
$^{(1)}$}
\centerline{T. Marshall Eubanks,$^{(3)}$, and Demetrios N. Matsakis,
$^{(3)}$}
\vskip .3cm
\centerline{\it $^{(1)}$Harvard-Smithsonian Center for Astrophysics}
\centerline{\it Cambridge, Massachusetts\quad 02138}
\centerline{\it email: pyne@cfa160.harvard.edu; birkinshaw@mb1.harvard.edu}
\vskip .3cm
\centerline{\it $^{(2)}$Physics Department,}
\centerline{\it University of Santa Barbara,}
\centerline{\it Santa Barbara, California 93106}
\centerline{\it email: cgwinn@condor.physics.ucsb.edu}
\vskip .3cm
\centerline{\it $^{(3)}$U.S. Naval Observatory,}
\centerline{\it Washington D.C. 20392}
\centerline{\it email:TME@usno01.usno.navy.mil; dnm@orion.usno.navy.mil}
\vskip .5cm

\noindent
{\bf Abstract}

Gravitational waves affect the observed direction of light from
distant sources. At telescopes, this change in direction appears
as periodic variations in the apparent positions of these sources
on the sky; that is, as proper motion. A wave of a given phase,
traveling in a given direction, produces a characteristic pattern
of proper motions over the sky. Comparison of observed proper
motions with this pattern serves to test for the presence of
gravitational waves. A stochastic background of waves induces
apparent proper motions with specific statistical properties,
and so, may also be sought. In this paper we consider
the effects of a cosmological background of
gravitational radiation
on astrometric observations. We derive an equation for the
time delay measured by two antennae observing the same source in
an Einstein-de Sitter spacetime containing gravitational radiation.
We also show how to obtain similar expressions for curved
Friedmann-Robertson-Walker spacetimes.


\vskip0.5truein
\noindent
{\bf 1. Introduction}

It is commonly agreed that gravitational waves,
predicted by Einstein's theory of general relativity
(Einstein 1916), must exist in our Universe. To date, however,
the only evidence for their presence is that the orbital decay rates
of certain binary pulsars appear consistent with the
predicted rates of energy loss from gravitational radiation
(Taylor 1992; Taylor and Weisberg 1989).
Nevertheless, gravitational waves arise generically
after the inflationary phase in inflationary
cosmologies (Rubakov, Sazhin, and Veryaskin 1982;
Fabbri and Pollock 1983; Abbott and Wise 1984)
and should be produced in a wide range of
physical situations at later times (Thorne 1987; Carr 1980).
For these reasons astrophysicists
are certain that gravitational wave astronomy, though difficult,
will be of enormous value in understanding our Universe.

The effects of gravitational radiation may be divided into
two categories. Direct effects physically
couple the energy density in the
waves to matter, causing, for instance, a bar to resonate.
Gravitational waves also affect the propagation of
radiation, causing a spacetime containing gravitational waves to
look different from one without. The very long wavelength
($\lambda >10^{-3}$ pc)
gravitational radiation which we focus on in this paper is best
searched for by examining its effects on the radiation we receive
from astrophysical sources. Such indirect
effects have been used successfully to constrain the fraction
of the energy density in our Universe which can be contained in
gravitational radiation of various wavelengths. Among these constraints
are $\Omega_{\rm GW}<10^{-4}$ at $\lambda \approx 1$ pc, from
pulsar timing (Romani and Taylor 1983; Taylor 1987);
$\Omega_{\rm GW}<0.04$ at $10\ {\rm pc}\le \lambda\le 10$ kpc, also
from pulsar timing (Taylor and Weisberg 1989);
$\Omega_{\rm GW}<10^{-4}$ at $\lambda \le 0.1$ kpc if the waves
existed during nucleosynthesis, from nucleosynthesis
constraints (Carr 1980);
$\Omega_{\rm GW}<10^{-8}$ or $10^{-3}$ at $\lambda > 1$ Mpc
if the waves did or did not exist, respectively, at recombination,
from microwave background anisotropy limits
(Linder 1988a); and $\Omega_{\rm GW}<10^{-3}$ for
 30 kpc $\le \lambda \le 300$ Mpc, from galaxy-galaxy $n$-point
correlation functions (Linder 1988b).

Recently Eubanks and Matsakis (1994) have
reported Very Long Baseline Interferometry (VLBI)
measurements that indicate quasars have a definite pattern
of apparent motions on the sky with root-mean-square (RMS) angular
velocity $\sim$ 20 $\mu$as ${\rm yr}^{-1}$.
The work of Linder (1988b) furnishes an estimate of the constraint
we can expect from data of this accuracy. Linder obtains an expression
for the mean-square angular deflection of light from cosmological
sources induced by gravitational waves. Dividing this expression
by the square of the wave period we gain an estimate for
the mean-square angular velocity of
sources at redshift $z$ in an Einstein-de Sitter spacetime
induced by waves with
energy density as a fraction of the closure density $\Omega_{\rm GW}$
given by $\langle \omega^2\rangle \approx \left( 1+( 1+z)^2
\right)H_o^2\Omega_{\rm GW}$ where $H_o$ is the Hubble constant
at $z=0$. Using $H_o=100h\ {\rm km\ s}^{-1}\ {\rm Mpc}^{-1}$
this tells us (taking $z=1$ as a representative redshift)
that the quasar motion data may be expected to
either constrain or detect gravitational waves at a level of
$\Omega_{\rm GW}= 0.04{\rm h}^{-2}$. This is competitive with the
pulsar timing limits, and should cover a much larger range of
gravitational wave wavelengths.


In order to
test the hypothesis that the quasar motions reported by
Eubanks and Matsakis (1994) are caused by a cosmologically significant
background of gravitational waves we need
a theoretical framework suitable for the
analysis of the effects of such waves
on VLBI measurements of distant sources. In a seminal work
on observations in cosmology, Kristian and Sachs (1965)
established a number of formulae relating the observed properties
of cosmological sources to the physical properties of the
spacetime in which they are observed. Their formula for the
proper motion distance could be used immediately to analyze the
system we are concerned with here except that their work utilizes
an expansion in the distance to the source divided by
some reasonably defined radius of curvature of the spacetime.
For the high redshift quasars of Eubanks and Matsakis (1994)
such an expansion is not useful as the quasars are a large fraction
of the Hubble distance, and many gravitational wave wavelengths,
from us.
The work of Linder (1988b) also bears close relationship
to the problem under consideration. Linder has obtained the deviation in
angle suffered by a light ray in an Einstein-de Sitter spacetime
containing gravitational waves. For astrophysical thin lens systems,
where the angular deflection of an incident light ray may be
considered to occur at a single point, knowledge of the angular
deviation is sufficient to determine the apparent position of the
source on the observer's sky. This is the content of the well known
lens equation. The situation is different, however, when a cosmological
background of gravitational waves is effectively acting as the lens.
For the wave case, the angular deflection occurs over the entire
photon path and there is no obvious {\it a priori} relationship
between the source position on the observer's sky and the
purely mathematical, i.e. unobservable, angular deviation.

In this paper we use the perturbative geodesic expansion introduced
in Pyne and Birkinshaw (1993) to determine the effects of
gravitational radiation on VLBI measurements of distant
sources. In
Gwinn {\it et al.} (1995) we use the results of this work to
test the hypothesis that the quasar motions reported by
Eubanks and Matsakis (1994) are caused by a cosmologically significant
background of gravitational waves.

The outline of this paper is as follows.
In section 2 we develop a method for analyzing a VLBI
experiment in metric perturbed Einstein de-Sitter
spacetimes. In section 3 we apply the method to the case
of an Einstein-de Sitter spacetime perturbed by a spectrum
of cosmological gravitational waves. In section 4 we consider
a single, plus-polarized, monochromatic wave and determine the
pattern of source proper motions which it produces on
the sky. In section 5 we
generalize our equations to the curved Friedmann-Robertson-Walker
(FRW) spacetimes and show how a physical understanding of the
method emerges from an analysis of the Jacobi equation.
In section 6 we investigate the consistency of our equations by
considering a simple gravitational lens system.
In section 7 we present
our conclusions.

\vskip0.5truein
\noindent
{\bf 2. Light Rays in a Perturbed Einstein-de Sitter Spacetime}

\nobreak
The results of Pyne and Birkinshaw (1995) allow construction of the
paths of light rays through a perturbed Einstein-de Sitter
spacetime with only minimal effort. The metric for such a
spacetime takes the form

$$d{\bar s}^2=a^2\left(-d\eta^2 +
 dx^2+dy^2+dz^2 \right)
+a^2h_{\mu\nu}dx^{\mu}dx^{\nu}  \eqno{(1)}  	$$

\noindent
where $a$ is the Friedmann expansion factor. We let
$d{\bar s}^2=a^2ds^2$.
By standard conformal results,
light rays in $d{\bar s}^2$ and $ds^2$ coincide and their
(affine) parameterizations are related by ${\bar
k}^{\mu}=a^{-2}k^{\mu}$. Here, and throughout this paper,
Roman letters $i,j,...$ run over
$\lbrace 1,2,3\rbrace$, while Greek letters $\mu ,\nu ,...$ run over
$\lbrace 0,1,2,3 \rbrace$. We use geometrized units, $G=c=1$.
The spacetime metric
is taken to have signature $+2$.
Our Riemann and Ricci tensor conventions are given by
$\left[ \Delta_{\alpha},\Delta_{\beta}\right] v^{\mu}=
R^{\mu}{}_{\nu\alpha\beta}v^{\nu}$ and $R_{\alpha\beta}=
R^{\mu}{}_{\alpha\mu\beta}$.

The metric, (1), is of a class of metrics whose radial null
geodesics were investigated in Pyne and Birkinshaw (1995). For the
specific case of (1), the Einstein-de Sitter background, non-radial
null geodesics may be constructed from the results of that paper
with almost no effort (this is because the Jacobi and parallel
propagators for these geodesics are the same as those used in that work.
For the case of
the curved FRW backgrounds, the propagators for the
radial and the non-radial geodesics are not
equivalent). We express this in the
form of a

\noindent
{\bf Theorem}: Let $ds^{(0)2}$ denote that part of
$ds^2$ independent of the perturbation (i.e. the Minkowski metric).
Let $x^{(0)\mu}(\lambda )$ be an affinely parametrized null
geodesic of $ds^{(0)2}$ with $k^{(0)\mu}(\lambda )=dx^{(0)\mu}(\lambda )
/d\lambda $ such that $k^{(0)0}=1$. Put

$$ f^{(1)\mu}=-\Gamma^{(1)\mu}{}_{\alpha\beta}k^{(0)\alpha}
k^{(0)\beta} .\eqno{(2)}$$

\noindent
with $\Gamma^{(1)\mu}{}_{\alpha\beta}$ that part of the Christoffel
connection of $ds^{2}$ which is linear in the metric
perturbation and its first partial derivatives. Then
$x^{\mu}(\lambda )=x^{(0)\mu}(\lambda )+x^{(1)\mu}(\lambda )$ is
an affinely parametrized geodesic (not necessarily null)
of $ds^2$ to first order provided that, for all $\lambda_2$,
$\lambda_1$

$$ x^{(1)\mu}\left( \lambda_2\right)=x^{(1)\mu}\left( \lambda_1\right)
+\left(\lambda_2 -\lambda_1\right)k^{(1)\mu}\left(\lambda_1\right)
+\int_{\lambda_1}^{\lambda_2} \left( \lambda_2 -\lambda\right)
f^{(1)\mu}(\lambda )\, d\lambda \eqno{(3)}$$

\noindent
where $k^{(1)\mu}(\lambda )=dx^{(1)\mu}(\lambda )/d\lambda $ and
the integration is performed over $x^{(0)\mu}(\lambda )$.

The condition $k^{(0)0}=1$ on the affine parametrization of the
background geodesic in the above theorem is imposed simply so that we
can use the propagators of Pyne and Birkinshaw (1995), who
imposed that condition for ease of calculation. It is not hard to
compute the necessary propagators for $k^{(0)\mu}(\tau )$ with
$\tau$ any affine parameter. This is not really needed, however,
since the geodesic, $k^{\mu}(\lambda )$,
constructed by the theorem above may be reparametrized directly.
For this reason we will sometimes refer to $k^{\mu}(\lambda )$ as
a wavevector though this term is usually reserved for the tangent
to a null geodesic parametrized so that $u_{\mu}k^{\mu}$ is the
photon frequency observed by an observer with four-velocity
$u^{\mu}$.

Suppose now that we solve (3) along some given
geodesic of $ds^{(0)2}$, $x^{(0)\mu}(\lambda )$
subject to $x^{(1)\mu}\left( \lambda_2\right)=0$ and
$x^{(1)i}\left( \lambda_1 \right)=0$. We can not simply
demand that the separation, $x^{(1)\mu}(\lambda )$,
vanish at both $\lambda_1$ and
$\lambda_2$ if we want the constructed geodesic to be null since
$x^{(0)\mu}\left( \lambda_1\right)$ and $x^{(0)\mu}\left( \lambda_2\right)$
are null separated in $ds^{(0)2}$ but not necessarily
in $ds^2$. Because the spatial and timelike components of (3)
decouple, however, we can use the above boundary conditions to
obtain $k^{(1)i}\left( \lambda_1\right)$. We can then
solve for $x^{(1)0}\left( \lambda_1\right)$ by demanding that
our constructed geodesic be null.

The condition that our constructed geodesic be null in
$ds^2$ can be written

$$k^{(1)0}=
{1\over 2}k^{(0)\mu}h_{\mu\nu}k^{(0)\nu}+k^{(0)i}\eta_{ij}k^{(1)j}
\eqno{(4)}$$

\noindent
where the equation holds along $x^{(0)\mu}\left( \lambda\right)$.
Taking our boundary conditions into account, this allows us to write the
timelike component of (3) as

$$\eqalign{ x^{(1)0}\left( \lambda_1\right) &=
-{1\over 2}\left( \lambda_2 -\lambda_1\right)k^{(0)\mu}
h_{\mu\nu}k^{(0)\nu} -\left( \lambda_2-\lambda_1 \right)
k^{(0)i}\eta_{ij}k^{(1)j} \cr & \qquad -
\int_{\lambda_1}^{\lambda_2}\left( \lambda_2 -\lambda\right)
f^{(1)0}(\lambda )\, d\lambda ,\cr} \eqno{(5)}$$

\noindent
the inner products being evaluated at
$x^{(0)\mu}\left( \lambda_1\right)$.
The spatial components of (3) yield

$$\left( \lambda_2 -\lambda_1\right)k^{(1)i}\left( \lambda_1\right)=
-\int^{\lambda_2}_{\lambda_1}
\left( \lambda_2 -\lambda \right)f^{(1)i}(\lambda )\, d\lambda \eqno{(6)}$$

\noindent
which may be combined with (5) to produce

$$x^{(1)0}\left( \lambda_1\right)=-{1 \over 2}
\left( \lambda_2 -\lambda_1\right)k^{(0)\mu}
h_{\mu\nu}k^{(0)\nu}+\int^{\lambda_2}_{\lambda_1}
\left( \lambda_2-\lambda \right)k^{(0)}_{\mu}f^{(1)\mu}
(\lambda )\, d\lambda \eqno{(7)}$$

\noindent
Equation (7) could also have been obtained immediately from (3)
and (4) after taking the inner product of (3) with $k^{(0)\mu}$.

At this point another representation of the perturbation vector,
$f^{(1)\mu}$, is very useful. Letting a semicolon denote covariant
differentiation using the Christoffel connection of $ds^{(0)2}$, we
have

$$f^{(1)\mu}={1\over 2}h_{\alpha\beta}{}^{;\mu}k^{(0)\alpha}k^{(0)\beta}
-h^{\mu}{}_{\alpha ;\beta}k^{(0)\alpha}k^{(0)\beta} .\eqno{(8)}$$

\noindent
Since $k^{(0)\mu}$ is geodesic, this gives

$$k^{(0)}_{\mu}f^{(1)\mu}=-{1\over 2}{d\over d\lambda}\left(
k^{(0)\mu}
h_{\mu\nu}k^{(0)\nu}\right) .\eqno{(9)}$$

\noindent
This may be substituted into (7) allowing an integration by parts to
produce

$$x^{(1)0}\left( \lambda_1\right)=-{1\over 2}\int^{\lambda_2}_{\lambda_1}
\left(k^{(0)\mu}
h_{\mu\nu}k^{(0)\nu}\right)
\, d\lambda \eqno{(10)}$$

\noindent
We have thus found the following

\noindent
{\bf Corollary}: Given two points,
$q\equiv x^{(0)\mu}\left( \lambda_2
\right)$ and $w\equiv x^{(0)\mu}\left( \lambda_1\right)$, connected
by a null geodesic of $ds^{(0)2}$, $x^{(0)\mu}(\lambda )$, with
$k^{(0)0}=1$, then to first order the points
$w^{\prime}\equiv x^{(0)\mu}\left(
\lambda_1 \right) +\left( x^{(1)0}\left(\lambda_1\right) ,0^i\right)$
and $q$ are null separated in $ds^2$ along some geodesic
$x^{\mu}$ provided $x^{(1)0}
\left( \lambda_1\right)$ obeys (10). Further, the tangent vector
to $x^{\mu}$ at $w^{\prime }$ is set by (4) and (6) above.

\noindent
In Section 3 we will show how this corollary may be used to
analyze the effects of an arbitrary
metric perturbation on VLBI observations in the metric (1).

We hasten to point out that the theorem, and so corollary, above
have no more content than a direct integration of the linearized
geodesic equations of our spacetime. We have presented this
information in this manner for two reasons. First,
as we will demonstrate in the next section, the above
corollary is specifically adapted to an analysis of VLBI
experiments in the spacetimes we are considering. Second,
the presentation above is organized so as to facilitate
the generalization to curved backgrounds described in section
5 below. The reader will note that for scalar perturbations in the
longitudinal gauge (see e.g. Mukhanov, Feldman, and Brandenberger
1992), the RHS of (10) is the Shapiro delay evaluated along the
background path. We will return to the relationship between
(10) and lens systems in section 6.

It remains for us to discuss the consistency criteria for the
manipulations leading to our corollary. We defer an examination
of this topic till section 6, contenting ourselves here with
an informal remark.
Imagine a geodesic of $ds^{(0)2}$, ${\tilde x}
^{(0)\mu}(\rho )$ which intersects $w^{\prime}$ at affine
parameter value $\lambda_1$ with
tangent vector ${\tilde k}^{(0)\mu}_{w^{\prime}}
=k^{(0)\mu}_w+k^{(1)\mu}_w$,
where a subscript
$w$ ($w^{\prime}$) denotes evaluation at $w$ ($w^{\prime}$)
and $k^{(1)\mu}_w$ is set by (4) and (6). We can use this
geodesic in (3) to construct a null geodesic in $ds^2$, ${\tilde x}^{\mu}
(\rho )$, obeying ${\tilde x}^{\mu}\left( \lambda_1\right)=w^{\prime}$
and ${\tilde k}^{\mu}_{w^{\prime}}={\tilde k}^{(0)\mu}_{w^{\prime}}$.
At a minimum, then, we would expect (10) to be a consistent
solution of the fixed endpoint problem
only if ${\tilde x}^{\mu}\left( \lambda_2\right)=q$. Roughly
speaking, we would expect this to hold if the gravitational
effects of the perturbation are similar on $x^{(0)\mu}$ and
${\tilde x}^{(0)\mu}$ for equal affine parameter values,
that is if $x$ and ${\tilde x}$ pass through ``sufficiently similar"
metrics.

\vskip0.5truein
\noindent
{\bf 3. Gravitational Radiation and VLBI}

\nobreak
In order to use (10) to investigate the effects of a gravitational
wave background on a VLBI experiment we first need the
appropriate form for the metric perturbation.
The plane, monochromatic wave solution in
the synchronous gauge to the perturbed Einstein's
equations associated with (1) can be written as the
real part of

$$\eqalign{ h_{00} &=0 \cr
 h_{0i} &=0 \cr h_{ij}&={a_o\over a}\left[ h_+\left( {\vec p}\right)
 \left( Re_+R^T\right)_{ij} +h_{\times}\left( {\vec p}\right)
\left( Re_{\times}R^T\right)_{ij} \right]
e^{i\left( {\vec p}\cdot {\vec x}-
p\eta \right)}\cr
&={a_o\over a}H_{ij}e^{i\left( {\vec p}\cdot {\vec x}-
p\eta \right)} \cr} \eqno{(11)}$$

\noindent
(Hawking, 1966), where $a_0$ is some fiducial value of $a$,
$h_+$ and $h_{\times}$ are complex valued functions containing the
amplitude and phase information, $e_+$ and $e_{\times}$ are
polarization matrices which we represent by

$$\eqalign{ e_+ &=\pmatrix{1 & 0 & 0 \cr 0 & -1 & 0\cr 0 & 0 & 0 \cr}
\cr e_{\times} &=\pmatrix{0 & 1 & 0 \cr 1 & 0 & 0 \cr 0 & 0 & 0 \cr}
\cr} ,\eqno{(12)}$$

\noindent
$R$ is a rotation matrix, $R^T$ its transpose, which we represent by

$$ R(\theta , \phi )=\pmatrix{ \sin\phi &\cos\theta\cos\phi &
\sin\theta\cos\phi \cr -\cos\phi & \cos\theta\sin\phi &
\sin\theta\sin\phi \cr 0 & -\sin\theta & \cos\theta \cr},\eqno{(13)}$$

\noindent
and the modevector,

$$\eqalign{ p^i&=p\pmatrix{ \sin\theta\cos\phi \cr \sin\theta\sin\phi \cr
\cos\theta \cr} \cr &=R\pmatrix{ 0\cr 0\cr p\cr} \cr}.\eqno{(14)}$$

\noindent
We can think of the angles, $\theta$ and $\phi$, as functions
of $\vec p$ through (14).
This solution is valid in what is known as the adiabatic regime,
which demands that, at the times of interest, the physical reduced
wavelength of the wave is much smaller than the Hubble distance.
Mathematically this is simply $1\gg a^{\prime}/p$ where ${}^
{\prime}\equiv d/dt$,
$t$ the comoving time co-ordinate related to $\eta$ by
$d/dt=a^{-1}d/d\eta$. The comoving reduced
wavelength, $1/p$, is related to the
physical wavelength on a spatial hypersurface of constant conformal
time $\eta$,
$\lambda_{\rm phys}$, by $\lambda_{\rm phys}=
2\pi a(\eta )/p$. The condition that a wave be in the adiabatic regime
at (unperturbed) redshift $z$ is thus
$\lambda_{\rm phys}\ll 2\pi (1+z)^{-3/2}H_o^{-1}$,
with $H_o$ the Hubble constant at $z=0$.
The general tensor
perturbation in the adiabatic regime, $h_{\mu\nu}\left(\eta ,
{\vec x}\right)$, is a superposition of the plane, monochromatic
waves for different modevectors.

We now show how to use the corollary of section 2 to analyze
a VLBI experiment in a metric perturbed Einstein-de Sitter spacetime
with perturbation given by (11). The basic idea is to construct
the future null cone of a point of emission by determining the
time at which the cone intersects any given line of constant
spatial position. This will allow us easily to determine the
difference in time of reception of a signal at two points a known
spatial distance apart, that is, at two ends of an interferometer.
The geometry of this section is shown
in Figure 1.

Suppose an antenna, which we label
antenna $A$, receives at the point $w_A^{\prime}\equiv
\left( \eta_A^{\prime}, {\vec r}_A\right)$ photons from a source
at $q\equiv \left( \eta_o -L, L{\hat e}_s\right)$.
We take ${\hat e}_s$ to be normalized
to unity in $ds^{(0)2}$. Denote the translation of
$w_A^{\prime}$ through time by
$\tau_A\equiv \left( \eta ,{\vec r}_A\right)$. The forward null cone
in $ds^{(0)2}$ of $q$ intersects $\tau_A$ at the point
$w_A=\left( \eta_o +\Delta_A ,{\vec r}_A\right)$ where

$$\Delta_A =\left| L{\hat e}_s-{\vec r}_A \right| -L .\eqno{(15)}$$

\noindent
In (15) and below, the $ds^{(0)2}$ inner product is written
using vertical bars. We will also use a raised dot for this
inner product. We note that $\Delta_A$ has a very simple interpretation:
two antennae, one at the spatial origin and one at spatial
co-ordinates given by ${\vec r}_A$, observing
a source at $q$ in an unperturbed Einstein-de Sitter spacetime
would measure a time delay of  $\Delta_A$ between the reception of
a signal at the origin at time $\eta_o$ and the reception of the
same signal by the antenna at ${\vec r}_A$.
The connecting null geodesic is given by
$x^{(0)\mu}_A(\lambda )=\left( \lambda ,
L{\hat e}_s-\left(\lambda -\eta_o +L\right){\hat e}_A\right)$ with

$${\hat e}_A={L{\hat e}_s-{\vec r}_A\over \left|
L{\hat e}_s-{\vec r}_A\right|} \eqno{(16)}$$

\noindent
which is seen to be properly normalized.
The corollary of section 2 now demands

$$\eta_A^{\prime} =\eta_o +\Delta_A
-{1\over 2}\int^{\eta_o -L}_{\eta_o +\Delta_A}
\left( k^{(0)\mu}
h_{\mu\nu}k^{(0)\nu}\right)
\, d\lambda \eqno{(17)}$$

\noindent
the integral being taken over $x^{(0)\mu}_A(\lambda )$.

It will be convenient for us to use not (17), but the equivalent expression
in terms of the comoving time. Provided that the
change in the scale factor over $\Delta_A$ and
$\eta^{(1)}_A$ may be neglected

$$\eqalign{ t_A^{\prime} &=t_o +a_o\Delta_A
+t^{(1)}_A \cr
&=t_o +a_o\Delta_A -{a_o\over 2}\int^{\eta_o-L}_{\eta_o +\Delta_A}
\left( k^{(0)\mu}
h_{\mu\nu}k^{(0)\nu}\right)
\, d\lambda ,\cr} \eqno{(18)}$$

\noindent
the integration occuring over $x^{(0)\mu}_A(\lambda )$.

We now consider a second antenna, antenna $B$, which receives
photons at $w^{\prime}_B\equiv \left( \eta_B , {\vec r}_B\right)$
from the same point of emission, $q$. This system obeys
equations (15) and (18) when all of the subscript $A$'s are
replaced by subscript $B$'s. The measured (comoving) time delay
for the two antenna system is given by

$$\eqalign{ T_d &=t_A^{\prime}-t_B^{\prime} \cr
&=a_o\left( \Delta_A -\Delta_B\right) +t^{(1)}_A-t^{(1)}_B \cr}
\eqno{(19)}$$

\noindent
with $t^{(1)}_A$ given by the term in (18) linear in
$h_{\mu\nu}$ and $t^{(1)}_B$ given by the analogous
expression for antenna $B$.

To see that (19) contains the usual expression for the
time delay of a VLBI system, consider the term,
$T_d^{(0)}$, in (19)
which is independent of the perturbation. Put ${\vec r}_B={\vec r}_A
+{\vec b}={\vec r}+{\vec b}$. We take
the distance to the source to be much larger than
the proper lengths of either
${\vec r}$ or ${\vec b}$, and we work
to first order in the implied small quantities, e.g ${\vec b}/L$.
This is the mathematical expression of the locally plane wave
approximation whereby we neglect the curvature of the wavefronts at
the observer. With this approximation we have ${\hat e}_s\approx
{\hat e}_A$ and

$$ a_o\left( \Delta_A -\Delta_B\right)
\approx a_o{\vec b}\cdot {\hat e}_s  \eqno{(20)}$$

\noindent
Continuing, we note that the scale factor
$a_o\approx a\left( t_A^{\prime}\right)\equiv a(t)$ where we
have written the time of reception at $A$, $t^{\prime}_A$, simply
as $t$. Provided antenna $B$ does not move significantly over
$T_d$ we can write ${\vec b}$ as ${\vec b}(t)$, the (comoving)
co-ordinate vector reaching from antenna $A$ to antenna $B$ at
time $t$. In addition, we can give meaning to ${\hat e}_s(t)$ as
follows: the spatial co-ordinates of the source at the event
of emission of the photons which arrive at antenna $A$ at time $t$
is given by $L(t){\hat e}_s(t)$, with ${\hat e}_s(t)$ normalized
to unity in $ds^{(0)2}$. The first term in (19) can then be written
$a{\vec b}\cdot {\hat e}_s $, with each quantity evaluated at $t$.

The tensor which projects the wavevector of the arriving photons
${\bar k}^{\mu}$ into the rest space of the observing antennae
is written $\delta^{\mu}_{\nu}+u^{\mu}u_{\nu}$, with
$u^{\mu}$ the antennae four-velocity. Taking our antennae comoving
(for simplicity, a Lorentz boost will easily yield the general
case from the specific) the properly normalized
(in $d{\bar s}^2$) four-velocity
is given by $u^{\mu}=(1/a ,{\vec 0})$. This writes the unit
(in $d{\bar s}^2$) vector in the rest space of the observer which
points toward the source, $s^{\mu}$, as

$$s^{\mu}=-{ {\bar k}^{\mu}\over \left| u{\bar \cdot}{\bar k}\right| }
-u^{\mu}{ \left( u{\bar \cdot}{\bar k}\right)\over
\left| u{\bar \cdot}{\bar k}\right|} .\eqno{(21)}$$

\noindent
It is easy to check that $s^{\mu}$ has no timelike component.
In (21) we have used ${\bar \cdot}$ to denote the ${\bar g}_{\mu\nu}$
inner product. Since ${\bar k}^{\mu}=a^{-2}k^{\mu}=
a^{-2}k^{(0)\mu}+a^{-2}k^{(1)\mu}$ we have

$$\eqalign{ as^i&={\hat e}^i-k^{(1)i}-k^{(1)0}{\hat e}^i \cr
&={\hat e}^i-k^{(1)i}-{1\over 2}{\hat e}^mh_{mn}{\hat e}^n{\hat e}^i
+{\hat e}_mk^{(1)m}{\hat e}^i \cr}\eqno{(22)}$$

\noindent
where we have used (4) and we have written ${\hat e}_A$ simply
as ${\hat e}$. At zeroth order this allows us to write (19) as

$$\eqalign{ T_d^{(0)}&=a_o^2b^i\eta_{ij}s^{(0)j} \cr
 &=\left( b^i{\bar g}_{ij}s^j\right)^{(0)} \cr}. \eqno{(23)}$$

\noindent
This formula, scalar under spatial transformations in the
antenna rest space and whose quantities are simply the restriction
to this space of tensor quantities (which by definition have
well defined properties under Lorentz transformation), is
exactly the standard formula of VLBI (Thompson, Moran,
and Swenson 1986).

Under the locally plane wave approximation,
$\left( b^i{\bar g}_{ij}s^j\right)$ is the spatial distance in the
observer's rest frame that the incoming wavefront must still travel to
antenna $B$ at the instant it has hit antenna $A$. It is, thus,
the time delay. In this approximation, then, we can obtain the
time delay either by determining the perturbed direction vector to
the emitter, $s^i$, and expanding its inner product with the
baseline vector, $b^i$, or we can work directly from
(19), imposing the mathematical constraints which lead to locally
plane waves at the observer. We now show that these methods
agree by calculating the explicit formula for the
time delay using both methods

We have already seen that the zeroth order expressions agree.
Unfortunately, the proof to first order is not so simple.
We start by setting $a_0=1$ in order to simplify our notation.
This means that at $w_A$ there is no distinction between
$d{\bar s}^2$ and $ds^2$. Since we will
only be concerned with quantities at $w_A$ in our formulae, we
will, for the rest of this section, no longer put a bar over the
physical metric. Nevertheless, its presence is implied, that is,
e.g. $h_{\mu\nu}$ may be understood as ${\bar h}_{\mu\nu}$, the two
being equal at the point we need them.

We first calculate the first order
term in the time delay from the
formula

$$T_d^{(1)}=\left( b^i g_{ij}s^j\right)^{(1)} \eqno{(24)}$$

\noindent
{}From (22) we see that we need the perturbation to the
spatial components of the wavevector at antenna $A$, $k^{(1)i}_A$.
This quantity is solved for
by (6), above, using the background path $x^{(0)\mu}_A$,
with the result

$$\eqalign{ k^{(1)i}_A &=\left( {i\over 2}
{\hat e}_A\cdot  H{\hat e}_A{\vec p}-ip\left( 1+{\hat p}\cdot
{\hat e}_A\right)H{\hat e}_A\right) {\eta_o^2\over L+\Delta_A}
e^{i{\vec p}\cdot {\vec r}} \cr
&\qquad \times e^{i{\vec p}\cdot {\hat e}_A\left( \eta_o +\Delta_A
\right) }J \cr}\eqno{(25)}$$

\noindent
with

$$J=\left( \eta_o -L\right)\int_{\eta_o+\Delta_A}^{\eta_o-L}
\lambda^{-2}
e^{-ip\left( 1+{\hat p}\cdot {\hat e}_A\right)\lambda}\, d\lambda
-\int_{\eta_o+\Delta_A}^{\eta_o-L}\lambda^{-1}
e^{-ip\left( 1+{\hat p}\cdot {\hat e}_A\right)\lambda}\, d\lambda
\eqno{(26)}$$

\noindent
The integrals may be performed explicitly, resulting in combinations
of exponential and exponential-integral functions of arguments
$-ip\left( 1+{\hat p}\cdot{\hat e}_A\right)\left( \eta_0 -L\right)$ and
$-ip\left(
1+{\hat p}\cdot{\hat e}_A\right)\left( \eta_0 +\Delta_A\right)$.
The result, while correct, is not particularly easy to use and we do
not write it down. Instead,
we note that adiabaticity demands $p\left( \eta_0 -L\right)
\gg 1$, which allows us to use the large argument expansion for
the exponential-integral functions (e.g. Gradshteyn and Ryzhik 1994,
equation 8.215) for all values of ${\hat e}_A$ not
too close to $-{\hat p}$. Next we suppose that our source is many
gravitational wave (reduced) wavelengths away, $pL\gg 1$,
and we agree to work only to leading order in $(pL)^{-1}$.
The solution, for the
appropriate range of ${\hat e}_A$, then becomes

$$\eqalign{ s^i& ={\hat e}^i_A+{\eta_0^2 \over \left( \eta_0
+\Delta_A\right)^2}
e^{i{\vec p}\cdot {\vec r}_A}e^{-ip\left( \eta_0 +\Delta_A\right)} \cr
 &\qquad \times
{{\hat e}_A\cdot H{\hat e}_A \over
2\left( 1+ {\hat p}\cdot{\hat e}_A\right)}\left( {\hat e}_A+ {\hat p}\right)
\ .\cr}\eqno{(27)}$$

\noindent
Because this expression shares with the exact result, (25)
and (26) above, the property
of vanishing at ${\hat e}_A=-{\hat p}$ we will use it for all
values of ${\hat e}_A$. Rigorously, we should examine its approach
to this zero and compare it with that of the exact expression, but
we will consider ourselves justified by the end result.
Noting that, to first order,

$$\eqalign{h_{ij}\left( w_A^{\prime}\right)&=h_{ij}
\left( w_A\right) \cr
&={\eta_0^2 \over \left( \eta_0
+\Delta_A\right)^2}
e^{i{\vec p}\cdot {\vec r}_A}e^{-ip\left( \eta_0 +\Delta_A\right)}
H_{ij} \ ,\cr}\eqno{(28)}$$

\noindent
equation (27) writes for the time delay

$$\eqalign{ T_d&=\left( b^ig_{ij}s^j\right) \cr
&={\vec b}\cdot {\hat e}_A+{1\over 2}{\hat e}_A^ih_{ij}{\hat e}_A^j
{{\vec b}\cdot {\hat e}_A+{\vec b}\cdot {\hat p}
\over \left(1+ {\hat p}\cdot{\hat e}_A\right)} \cr}\eqno{(29)}$$

\noindent
We note that this formula is a well-defined scalar in
rest space of our observer. We also point out that ${\hat p}=
{\hat p}_{\rm phys}$ and that our hatted vectors are of unit norm
in the physical background metric at the observer since,
briefly returning to our old bar notation, ${\bar g}^{(0)}_{\mu\nu}
=\eta_{\mu\nu}$. Defining the dimensionless vector
${\vec \zeta}={\hat e}_A+{\hat p}$, with ${\hat \zeta}\equiv
{\vec \zeta}/\sqrt{ \zeta^i\eta_{ij}\zeta^j} $,
we have

$$T_d={\vec b}\cdot {\hat e}_A+{\vec b}\cdot {\vec \zeta}
\left( {\hat \zeta}^ih_{ij}{\hat \zeta}^j
\right)  .\eqno{(30)}$$

\noindent
In the following section we will show how this formula may be used
to obtain the pattern of proper motions on the sky measured
by an interferometer observing
a sample of distant sources through
a background of gravitational radiation.

For completeness, we now show how our solution, (30),
is obtained from (18) and
(19), above. These write the first order contribution
to the time delay as

$$\eqalign{ T_d^{(1)} &=t^{(1)}_A-t^{(1)}_B \cr
  &=-{1\over 2}\int^{\eta_o-L}_{\eta_o +\Delta_A}
k^{(0)\mu}_A
h_{\mu\nu}\left( x^{(0)}_A\right)k^{(0)\nu}_A
\, d\lambda +
{1\over 2}\int^{\eta_o-L}_{\eta_o +\Delta_B}
k^{(0)\mu}_B
h_{\mu\nu}\left( x^{(0)}_B\right)k^{(0)\nu}_B
\, d\lambda .\cr} \eqno{(31)}$$

\noindent
In this formula, the path on which $h_{\mu\nu}$ is to be evaluated
has been shown explicitly.

The easiest way to proceed is to expand the integrand
in the integral for $t^{(1)}_B$ about the value of the integrand in
the integral for $t^{(1)}_A$ for fixed $\lambda$. If we neglect terms
of quadratic order in ${ \vec b}/L$,

$$\eqalign{ k^{(0)\mu}_Bh_{\mu\nu}\left( x^{(0)}_B\right)k^{(0)\nu}_B &=
k^{(0)\mu}_Ah_{\mu\nu}\left( x^{(0)}_A\right)k^{(0)\nu}_A +
2\delta k^{(0)\mu}h_{\mu\nu}\left( x^{(0)}_A\right)k^{(0)\nu}_A \cr
&\qquad +
k^{(0)\mu}_Ah_{\mu\nu ,\alpha}\left( x^{(0)}_A\right)
\delta x^{(0)\alpha}k^{(0)\nu}_A \cr} \eqno{(32)}$$

\noindent
with

$$\eqalign{ \delta k^{(0)\mu}&=k^{(0)\mu}_B-k^{(0)\mu}_A \cr
&={1\over L}\pmatrix{ 0\cr {\vec b}-( {\vec b}\cdot {\hat e}_s )
{\hat e}_s \cr} \cr} \eqno{(33)}$$

\noindent
and $\delta x^{(0)\mu}(\lambda )=\left( \lambda +L-\eta_0\right)
\delta k^{(0)\mu}$. Using (32) in (31) yields

$$\eqalign{ T_d^{(1)} &={1\over 2}\int^{\eta_0 -L}_{\eta_0 +\Delta_A}
2\delta k^{(0)\mu}h_{\mu\nu}\left( x^{(0)}_A\right)k^{(0)\nu}_A +
k^{(0)\mu}_Ah_{\mu\nu ,\alpha}\left( x^{(0)}_A\right)
\delta x^{(0)\alpha}k^{(0)\nu}_A \, d\lambda \cr
&\qquad +{1\over 2}
k^{(0)\mu}_Ah_{\mu\nu}\left( p_A\right)k^{(0)\nu}_A \cr
&={1\over 2}
k^{(0)\mu}_Ah_{\mu\nu}\left( p_A\right)k^{(0)\nu}_A \cr
&\qquad -{\hat e}_A\cdot H\left( {\vec b}-({\vec b}\cdot {\hat e}_s)
{\hat e}_s\right)
{\eta_0^2\over L} e^{i{\vec p}\cdot {\vec r}_A}e^{-i{\vec p}
\cdot {\hat e}_A\left( \eta_0 +\Delta_A\right)}I \cr
&\qquad -{i\over 2}{\hat e}_A\cdot H{\hat e}_A {\vec p}\cdot
\left( {\vec b}-({\vec b}\cdot {\hat e}_s)
{\hat e}_s\right)
{\eta_0^2\over L} e^{i{\vec p}\cdot {\vec r}_A}e^{-i{\vec p}
\cdot {\hat e}_A\left( \eta_0 +\Delta_A\right)}J \cr}\eqno{(34)}$$

\noindent
where $J$ is given by (26)  and

$$I=\int_{\eta_o+\Delta_A}^{\eta_o-L}\lambda^{-2}
e^{-ip\left( 1+{\hat p}\cdot {\hat e}_A\right)\lambda}\, d\lambda.
\eqno{(35)}$$

Once again, the integrals we need to perform yield a combination
of exponential and exponential-integral functions of
arguments $-ip\left( 1+{\hat p}\cdot{\hat e}_A\right)
\left( \eta_0 -L\right)$ and $-ip\left(
1+{\hat p}\cdot{\hat e}_A\right)\left( \eta_0 +\Delta_A\right)$.
We follow our earlier course and use the large argument
expansion of the exponential-integral functions. The terms in the
resultant expression contributed by the term written proportional
to $I$ in (34), are of order $1/pL$ times the terms contributed by
the term written proportional to $J$ in (34), and of order
$1/pL$ times the first term in (34) (in fact, direct dimensional
analysis of (34) argues for this conclusion). Again restricting ourselves
to sources which are many gravitational wave reduced wavelengths away,
we may neglect the terms down by $pL$.
The remaining terms combine easily to give the first order part of
(29), which is what we needed to show. That is, we have proven,
through first order,

$$ t_A-t_B=b^ig_{ij}s^j \eqno{(36)} $$

\noindent
with the LHS assembled from (15), (18), and (19),
and the RHS constructed from (22).

Because this has been a long section we will collect the
approximations used to gain our formula here. We envision an
adiabatic background of gravitational radiation and two
antennae, $A$ and $B$, such that the co-ordinate separation
vector, ${\vec b}$, reaching from antenna $A$ to antenna $B$
has a proper length much smaller than the (reduced)
wavelengths of the waves,
$p{\vec b}\ll 1$. We suppose the antennae to be nearly comoving
over the timescale of the measured delay. We further suppose
that the sources observed are many gravitational wave
(reduced) wavelengths
away, $pL\gg 1$.
Provided that these inequalities hold, the measured time delay,
$T_d$, is given by (29), with ${\hat e}_A$ related to
the observed direction vector to the source, $s^i$, by (22)
(using (25) as well).
We note that all of our conditions may be relaxed if necessary,
by using the more precise equations (19) and (34).

We can obtain a very rough {\it a posteriori} consistency check on
the accuracy of our perturbative expansion in the following way.
We approximate the separation between the solution geodesic,
$x^{\mu}(\lambda )$, and $x^{(0)\mu}(\lambda )$ by
$Lk^{(1)}\left( \lambda_1\right)\sim Lh$, where $h$ is some
characteristic element of $h_{\mu\nu}$. That is, the separation
is roughly the distance to
the source times the angle between the two geodesics at the observer.
We must demand that the perturbations ``felt" by these two
geodesics are nearly identical, which means,
since $h_{\mu\nu}$ varies over scales larger than $1/p$, $Lh\ll p^{-1}$.
The idea is now to square this inequality and relate $h^2p^2$ to
the fraction of the closure density contained in the gravitational
radiation. An explicit model for the wave background is necessary
for this step. For illustration, consider a
stochastic background of gravitational radiation.
Then $\Omega_{\rm GW}\sim p^2h^2/H_0^2$. Since

$$L=2H_0^{-1}\left( 1-\left( 1+z\right)^{-1/2}\right) \eqno{(37)}$$

\noindent
we can write our constraint as

$$\Omega_{\rm GW}\ll {1\over 4}\left( 1-\left( 1+z\right)^{-1/2}
\right)^{-2} \eqno{(38)}$$

\noindent
This is to be interpreted as follows. If we use the formulae above to
infer, from sources at characteristic redshift $z$, the
presence of a stochastic background of gravitational waves with
an energy density of $\Omega_{\rm GW}$, then
a crude estimate of
consistency is furnished by the degree which (38) is found to
hold. We point out that for $z\sim 1$ this is not a
highly restrictive condition.

\vskip0.5truein
\noindent
{\bf 4. The Pattern of Proper Motions on the Sky}

\nobreak
In this section we will show how equation (30)
may be used to gain the pattern of proper motions
inferred by an interferometrist observing
distant sources through a background of gravitational radiation.
Since we are working in the linear regime, without loss of
generality we consider
the effect of a single plus-polarized monochromatic wave.
The effect of a general wave background may be found from the
results of this section using superposition. We choose
our coordinates such that the $z$-axis is aligned with the waves
direction of propagation. The modevector, $\vec p$,
is then given by ${\vec p}=p{\hat z}$ and the metric perturbation is
written

$$\eqalign{ h_{00} &=0 \cr
 h_{0i} &=0 \cr h_{ij}&={1\over a} h_+e^{ip\left( z-
\eta \right)}
 \left( e_+\right)_{ij}
\cr}. \eqno{(39)}$$

\noindent
As in the last section we have set $a_0=1$. Unlike there, however,
in this section we return to our practice of using an overbar to
distinguish between the perturbed Einstein-de Sitter spacetime
and its conformal relative, shown in (1).

The approximations of the previous section guarantee the existence
of a region of spacetime $\Sigma$ such that $\Sigma$ contains
the events of reception, $a\approx 1$ in $\Sigma$, and
$px^i\ll 1$ for any $x^{\mu} \in \Sigma$. This last condition simply states
that the spatial extent of $\Sigma$ on any constant conformal time
hypersurface is much smaller than the wavelength of the gravitational
wave. In $\Sigma$, to within our level
of approximation, the coordinate transformation

$$\eqalign{ t^{\prime}&=t \cr
x^{\prime}&=\left( 1+{1\over 2}h_+e^{ip\left(z-\eta
\right)}\right)x \cr
y^{\prime}&=\left( 1-{1\over 2}h_+e^{ip\left(z-\eta
\right)}\right)y \cr
z^{\prime}&=z \cr} \eqno{(40)}$$

\noindent
brings the actual metric to Minkowski form; ${\bar g}_{\mu^{\prime}
\nu^{\prime}}=\eta_{\mu^{\prime}
\nu^{\prime}}$. We note that in (40) $\eta$ is to be
considered a function of $t$ in the usual way.

For simplicity we again suppose at least one of the
antennae is comoving, having four velocity
$u^{\mu}=(1,0,0,0)$.
The surfaces of constant conformal
time then serve as instantaneous rest three-spaces for
an observer at this antenna.
The co-ordinate transformation (40) may be
considered a transformation of co-ordinates in these hypersurfaces.
Since the spatial components of $s^i$, given by (21) above,
define a vector in these hypersurfaces, the observed
direction to a source in the primed coordinates is simply

$$s^{i^{\prime}}={\partial x^{i^{\prime}}\over \partial x^i}
s^i \ ,\eqno{(41)}$$

\noindent
and the time delay measured by an antenna pair with separation
$b^{i^{\prime}}$ is given by

$$T_d=b^{i^{\prime}}\eta_{i^{\prime}j^{\prime}}s^{i^{\prime}}\ .
\eqno{(42)}$$

Consider now the triad of vectors

$$\eqalign{b_{(1)}&=B\left( 1-{1\over 2}h_+e^{-ip\eta}
\right){\hat x} \cr
b_{(2)}&=B\left( 1+{1\over 2}h_+e^{-ip\eta}
\right){\hat y} \cr
b_{(3)}&=B{\hat z} \cr}\eqno{(43)}$$

\noindent
with ${\hat x}$, ${\hat y}$, and ${\hat z}$ the unprimed
(spatial) co-ordinate
basis vectors (so, for example, ${\hat y}^{\mu}=(0,0,1,0)$ ).
The vectors
$b_{(i)}$ are orthogonal and have proper length
$B$ with respect to ${\bar g}_{\mu\nu}$. Their spatial components define
a triad of three-vectors in the observers rest space, orthogonal
and of proper length $B$ with respect to the induced metric
${\bar g}_{ij}$.
In terms of the primed (spatial) co-ordinate
basis vectors, ${\hat x}^{\prime}$, ${\hat y}^{\prime}$, and
${\hat z}^{\prime}$, the triad is written

$$\eqalign{b_{(1)}&=B{\hat x}^{\prime} \cr
b_{(2)}&=B{\hat y}^{\prime} \cr
b_{(3)}&=B{\hat z}^{\prime} \cr}\eqno{(44)}$$

We now suppose that the observer makes time delay measurements
of a source with ${\hat e}_A=\alpha {\hat x}+\beta {\hat y}
+\gamma {\hat z}$
using each of the three-vectors of the triad as a baseline.
Then (30), (39), and (43)
tell us that $T_{d(i)}$, the time delay measured along baseline
${\vec b}_{(i)}$, is given by

$$\eqalign{ T_{d(1)}&=\alpha B\left( 1-{1\over 2}h_+e^{-ip\eta}
\right)\left[ 1+{1\over 2}h_+ e^{-ip\eta}\left( {\alpha^2-
\beta^2  \over 1+\gamma}\right)\right] \cr
T_{d(2)}&=\beta B\left( 1+{1\over 2}h_+e^{-ip\eta}
\right)\left[ 1+{1\over 2}h_+ e^{-ip\eta}\left( {\alpha^2-
\beta^2  \over 1+\gamma}\right)\right] \cr
T_{d(3)}&=B\left[ \gamma +{1\over 2}h_+ e^{-ip\eta}\left( \alpha^2-
\beta^2  \right)\right] \cr} .\eqno{(45)}$$

\noindent
Since the time delay is a scalar under co-ordinate transformations in the
observer's rest space we can also compute the delays
in the primed co-ordinates,

$$ T_{d(i)}=b_{(i)}^{m^{\prime}}\eta_{m^{\prime}
n^{\prime}}s^{n^{\prime}} .\eqno{(46)}$$

\noindent
Using (44) in (46) and equating the resulting expression with
the right-hand side of (45) allows us to read off
the components of the normalized direction vector to the source in
the primed co-ordinates,

$$\eqalign{ s^{x^{\prime}}&=\alpha \left( 1-{1\over 2}h_+e^{-ip\eta}
\right)\left[ 1+{1\over 2}h_+ e^{-ip\eta}\left( {\alpha^2-
\beta^2  \over 1+\gamma}\right)\right] \cr
s^{y^{\prime}}&=\beta \left( 1+{1\over 2}h_+e^{-ip\eta}
\right)\left[ 1+{1\over 2}h_+ e^{-ip\eta}\left( {\alpha^2-
\beta^2  \over 1+\gamma}\right)\right] \cr
s^{z^{\prime}}&=\left[ \gamma +{1\over 2}h_+ e^{-ip\eta}\left( \alpha^2-
\beta^2  \right)\right] \cr} .\eqno{(47)}$$

The proper motion inferred by the observer is given by
$\mu^i=P^{i^{\prime}}{}_{\beta^{\prime}}
u^{\alpha^{\prime}}
s^{\beta^{\prime}}{}_{|\alpha^{\prime}}$
where $P^{\alpha^{\prime}}{}_{\beta^{\prime}}=
\delta^{\alpha^{\prime}}_{\beta^{\prime}}+u^{\alpha^{\prime}}
u_{\beta^{\prime}}$ is the
projector into the observer's rest space and a slash denotes
the covariant
derivative of the perturbed metric, $d{\bar s}^2$.
To first order this results in the simple $\mu^{i^{\prime}}=
s^{i^{\prime}}{}_{,t^{\prime}}$. A comoving source will have
$\alpha$, $\beta$, and $\gamma$ constant. Then

$$\eqalign{ {\vec \mu}&={ip\alpha\over 2} h_+e^{-ip\eta}
\left( 1- {\alpha^2-
\beta^2  \over 1+\gamma}\right){\hat x}^{\prime} \cr
&\qquad-{ip\beta\over 2} h_+e^{-ip\eta}
\left( 1+ {\alpha^2-
\beta^2  \over 1+\gamma}\right){\hat y}^{\prime} \cr
&\qquad-{ip\over 2} h_+e^{-ip\eta}
\left( \alpha^2-
\beta^2  \right){\hat z}^{\prime} \cr} .\eqno{(48)}$$

\noindent
Allowing the direction cosines to depend on time enables the effect
of source peculiar motion to be calculated. To the level of
accuracy considered here the effect of source motion is to
superpose the standard FRW proper motion results on the
gravitational wave pattern given by (48) (see e.g. Weinberg 1972,
Chapter 15).

The expression, (48), for the proper motion contains the
direction cosines $\alpha$, $\beta$, and $\gamma$ which are the
components of the unobservable vector ${\hat e}_A$ in the unprimed
co-ordinates. These differ from the direction cosines in the
primed co-ordinates, $\alpha^{\prime}$, $\beta^{\prime}$, and
$\gamma^{\prime}$ where ${\hat e}_A=\alpha^{\prime}{\hat x}^{\prime}
+\beta^{\prime}{\hat y}^{\prime}+\gamma^{\prime}{\hat z}^{\prime}$,
by terms of first order. Further, as can be seen from (47),
$\alpha$, $\beta$, and $\gamma$ describe the components of the
observed direction vector to the quasar, $s$, to zeroth order in the
primed (and unprimed) co-ordinates. Since ${\vec \mu}$ is entirely
first order we may reinterpret $\alpha$, $\beta$ and
$\gamma$ in (48) as the direction cosines to the source
in the primed co-ordinates, the difference between these quantities
being first order and so resulting only in an ignorable second
order correction to $\vec \mu$. In terms of the spherical polar
co-ordinates of the source in the primed frame, defined by $s=
\sin\theta^{\prime}\cos\phi^{\prime}{\hat x}^{\prime}+
\sin\theta^{\prime}\sin\phi^{\prime}{\hat y}^{\prime}+
\cos\theta^{\prime}{\hat z}^{\prime}$, we have

$$\eqalign{ \alpha &\approx\sin\theta^{\prime}\cos\phi^{\prime} \cr
\beta &\approx\sin\theta^{\prime}\sin\phi^{\prime} \cr
\gamma &\approx\cos\theta^{\prime} \cr}\eqno{(49)}$$

\noindent
the approximate sign denoting equality to zeroth order. Substitution
of (49) into (48) and resolution of the resultant expression into
components along the spherical polar basis vectors given by

$$\eqalign{{\hat \theta}^{\prime} &=\cos\theta^{\prime}\cos\phi^{\prime}
{\hat x}^{\prime}+\cos\theta^{\prime}\sin\phi^{\prime}{\hat y}^{\prime}
-\sin\theta^{\prime}{\hat z}^{\prime} \cr
{\hat \phi}^{\prime} &=-\sin\phi^{\prime}{\hat x}^{\prime}
+\cos\phi^{\prime}{\hat y}^{\prime} \cr}\eqno{(50)}$$

\noindent
produces the result

$${\vec \mu}={ip\over 2} h_+e^{-ip\eta}\sin\theta^{\prime}
\left( \cos 2\phi^{\prime}\, {\hat \theta}^{\prime}-\sin
2\phi^{\prime}\, {\hat \phi}^{\prime}\right) \eqno{(51)}$$

\noindent
expressing the inferred proper motion in terms of the wave parameters,
$p$ and $h_+$, and the angular co-ordinates of the source in
a gaussian normal frame (to the necessary order of approximation)
at the observer.

\vskip0.5truein
\noindent
{\bf 5. Generalization to Curved Backgrounds}

\nobreak
In this section, we will consider the general context for the calculations
of the preceeding sections. By using the Jacobi equation
of the background spacetime, we will be able to obtain a physical
description of the manipulations involved. We begin by
recalling that the perturbative geodesic expansion (PGE) constructs an
affinely parametrized geodesic of a general perturbed metric,
$g^{(0)}_{\mu\nu}+h_{\mu\nu}$, from an affinely parametrized
geodesic, $x^{(0)\mu}(\lambda )$, of the background
metric, $g^{(0)}_{\mu\nu}$,
by writing the sought after
geodesic as $x^{\mu}(\lambda )=x^{(0)\mu}(\lambda )+
x^{(1)\mu}(\lambda )$ and solving for $x^{(1)\mu}(\lambda )$, the
separation (Pyne and Birkinshaw 1993).
Let the Jacobi propagator along $x^{(0)\mu}(\lambda )$
be written in terms of its $4\times 4$ subblocks as

$$ U\left( \lambda_1 ,\lambda_2\right)=\pmatrix{
 A\left( \lambda_1 ,\lambda_2\right) & B\left( \lambda_1 ,\lambda_2\right)
\cr  C\left( \lambda_1 ,\lambda_2\right) &
 D\left( \lambda_1 ,\lambda_2\right) \cr}.\eqno{(52)}$$

\noindent
Then the equation governing
the separation is written

$$\eqalign{ P\left( \lambda_1 ,\lambda_2\right)x^{(1)}\left( \lambda_2
\right) &=A\left( \lambda_1 ,\lambda_2\right)x^{(1)}\left( \lambda_1
\right) +B\left( \lambda_1 ,\lambda_2\right)k^{(1)}\left( \lambda_1
\right) \cr
&\qquad +B\left( \lambda_1 ,\lambda_2\right)
\left[ {d\over d\lambda }P\left( \lambda_1 ,\lambda\right)
\right]_{\lambda=\lambda_1}x^{(1)}\left( \lambda_1\right) \cr
&\qquad +\int^{\lambda_2}_{\lambda_1}B\left( \lambda_2 ,\lambda\right)
 P\left( \lambda_1 ,\lambda\right)f^{(1)}\left( \lambda
\right)\, d\lambda \cr} \eqno{(53)}$$

\noindent
where $k^{(1)}=dx^{(1)}/d\lambda$, $P$ is the parallel propagator
along $x^{(0)\mu}(\lambda )$, and $f^{(1)}$ is the perturbation vector
given by (8), above. In (53),
we have employed a $4\times 4$ matrix notation so that, for example,

$$ P\left( \lambda_1 ,\lambda\right)f^{(1)}\left( \lambda
\right)\equiv P\left( \lambda_1 ,\lambda\right)^{\mu}{}_{\nu}
f^{(1)\nu}\left( \lambda
\right)\ . \eqno{(54)}$$

\noindent
Specific forms for the Jacobi and parallel propagators of the
curved FRW spacetimes can be found in Pyne and Birkinshaw (1995).

Equation (53) is the generalization of equation (3) of section 2.
The proper generalization of equation (4) of section 2,
the condition that $k^{\mu}(\lambda )$
be null to first order in the perturbed metric, is written

$$h_{\mu\nu}k^{(0)\mu}k^{(0)\nu}+2g^{(0)}_{\mu\nu}k^{(0)\mu}k^{(1)\nu}
+2g^{(0)}_{\mu\nu , \rho}x^{(1)\rho}k^{(0)\mu}k^{(0)\nu}=0 ,\eqno{(55)}$$

\noindent
the constraint holding at each point of $x^{(0)\mu}(\lambda )$ provided
it holds at any given point. In principle, equations (53) and (55)
allow us to carry out analysis of the geodesic problem
subject to fixed-endpoint (or mixed) boundary conditions
in arbitrary metric perturbed spacetimes.
Such boundary conditions are more applicable to certain
astrophysical systems, such as
multiple image lensing or VLBI, than the more common initial-value
boundary data which is useful, for instance, in studies of the
Sachs-Wolfe effect (Sachs and Wolfe 1967).

The primary use we will make of equations (53) and (55) in this paper,
however, is to understand the manipulations of sections 2 and 3.
The physical picture is made clear by recognizing that (53)
is exactly the Jacobi equation of the background spacetime subject
to a forcing perturbation $f^{(1)}$. Consider the fixed endpoint
solution between $q$ and $w_A^{\prime}$ which made use of the
unperturbed geodesic $x^{(0)\mu}(\lambda )$ between $q$ and
$w_A$. We will simply state the conclusion; the reader can carry
out the
steps explicitly and compare them with the development in
sections 2 and 3 to confirm the conclusion that we offer here.
The solution is constructed
in the following way. First, solve, using (53), for the spatial separation,
$\delta x\left( \lambda_2\right)$, on the constant
conformal time hypersurface containing
the point of emission, $q$,
attained by a perturbed null geodesic
which intersects $w_A$ with wavevector coincident with
$k^{(0)\mu}\left(\lambda_1\right)$.
We can think of this as tracing the photon into the past and determining
its intersection with the constant conformal time
hypersurface containing $q$. The proper pertubation to the spatial
components of the wavevector
at $w_A$, $k^{(1)i}\left(\lambda_1\right)$, is then taken to be
that perturbation at $w_A$ which, when considered
as an impulsive perturbation in the background Jacobi equation,
produces a deviation
vector on the constant conformal time hypersurface
containing $q$ equal to $\delta x\left( \lambda_2\right)$.
Speaking in informal language borrowed from gravitational lens theory,
we solve a standard initial-data problem to gain a spatial separation
in the source plane, then we convert that to an angle at the observer
using the angular diameter distance of the background.

{}From the spatial components, $k^{(1)i}\left(\lambda_1\right)$, we
determine the timelike component of the wavevector perturbation,
$k^{(1)0}\left( \lambda_1\right)$, by
imposition of the null constraint (55).
Imagine we now constructed a null geodesic of $g^{(0)}_{\mu\nu}$
which intersected $w_A$ with wavevector
$k^{(0)\mu}\left( \lambda_1\right)+k^{(1)\mu}\left( \lambda_1\right)$
and used this geodesic to obtain a perturbed geodesic which
intersects $w_A$ with coincident wavevector. Modulo questions of
consistency, which we will discuss below, the resultant
perturbed geodesic will intersect a time translate of $q$ but not,
generally, $q$ itself. That is, it will ``hit" the proper spatial
co-ordinates but will ``miss" the point $q$ by some offset in time.
We correct for this by ``moving" the point $w_A$ in time to another
point $w_A^{\prime}$. The necessary time translation is
$x^{(1)0}\left( \lambda_1\right)$.

It is clear that great simplifications have
resulted from the foliation of our
background spacetime by spatial hypersurfaces. While the FRW
backgrounds possess such a foliation a general background does not,
and we have not considered fixed-endpoint problems in such spacetimes.
In the next section we return to general questions of consistency,
using as a theoretical laboratory the classic astrophysical
example of a fixed-endpoint solution, the gravitational lens.

\vskip0.5truein
\noindent
{\bf 6. A Simple Gravitational Lens and Consistency}

\nobreak
We work in a Minkowski space background.
We consider an observer at the spatial origin of co-ordinates,
a lens, of mass $m$, located at spatial co-ordinates
$L{\hat x}$, and an emitter which emits a burst of photons at
spacetime event $q=( -2L,2L,0,0)$. Figure 2 illustrates
our geometry. The question we want to
answer is where on the sky does the observer see the emitter?
The answer, of course, is that the observer sees a circular ring
around the lens, with the angle between lens and ring given by the
Einstein angle, $\theta_E=\sqrt{2m/L} $
(in our example the ring appears only
for an instant, but this is unimportant). We want to see how this
result emerges from the PGE. We will not actually perform
any calculations in this section. Rather we will utilize the
description of the fixed endpoint solution in the above section
in conjunction with well known results and certain ideas of
Pyne and Birkinshaw (1993) to arrive at a
plausible understanding.

Briefly, we set up the mathematics of our lens system.
For a Minkowski space background with metric $\eta_{\mu\nu}=
{\rm diag}(-1,1,1,1)$, equation (53) takes the form

$$x^{(1)\mu}\left( \lambda_2\right)=x^{(1)\mu}\left( \lambda_1\right)
+\left( \lambda_2-\lambda_1\right) k^{(1)\mu}\left( \lambda_1\right)
+\int^{\lambda_2}_{\lambda_1}\left( \lambda_2-\lambda\right)
f^{(1)\mu}(\lambda )\, d\lambda \eqno{(56)}$$

\noindent
An appropriate perturbation for our lens can be determined in the
well known weak field approximation (see, e.g. Weinberg 1972,
Chapter 10) with the
result $h_{\mu\nu}={\rm diag}(-2\phi ,-2\phi ,-2\phi ,-2\phi )$
with $\phi$ the Newtonian potential of the lens on the spatial
hypersurfaces. It is easiest to suppose that both the emitter and the
observer are sufficiently far from the lens that its
gravitational field upon them may be neglected. For simplicity
we could also assume that both the emitter and the observer
have four-velocities given by $u^{\mu}=(1,0,0,0)$.

Consider the one parameter family of null geodesics of our Minkowski
space background given by

$$x^{(0)\mu}(\lambda )=\pmatrix{ \lambda \cr
2L-2L\cos\theta -\lambda\cos\theta \cr 2L\sin\theta +\lambda\sin
\theta \cr 0}\eqno{(57)}$$

\noindent
At this point, there is nothing mathematically to stop us from
constructing one null geodesic in the
perturbed spacetime from each of these background geodesics, subject
to the boundary conditions $x^{(1)\mu}(-2L )=0$ and
$x^{(1)i}( 0)=x^{(0)i}(0)$. In addition, we could find
$x^{(1)0}(0)$ by forcing
our constructed geodesics to be null in $\eta_{\mu\nu}+h_{\mu\nu}$.
These boundary conditions would ensure that each of the constructed
null geodesics intercepts both $q$ and the worldline of the
observer. Finally, for each perturbed geodesic we could
determine the angle it defined at the observer with the
image of the lens.

Of course, our ability to construct so many such geodesics is
an enormous warning sign: we know very well that at most two
can be good approximations to the exact solution, one passing
to each side of the lens.
How then can we single out the two good approximate geodesics
from the many bad ones? In Figure 2, we show three of the
background paths we are considering, projected into the
$xy$-plane. One, $x^{(0)\mu}_C$, travels diametrically
away from the lens, one, $x^{(0)\mu}_A$, travels extremely close to the
Schwarschild radius of the lens, $R_s$,
and one is that background path which generates the best
approximate perturbed path. The best approximation
background path, $x^{(0)\mu}_B$, defines an angle with the
$x$-axis at the emitter equal to $\theta_E$. If we take seriously
the cinematic description of our method offered in section 5,
we should demand that any background path be suitable
for an initial value calculation with initial value specified
at the observer. This would immediately rule out the use of
the path which passes very close to the lens (Pyne and
Birkinshaw 1993). The other extreme path, however, is perfectly
suitable for such a calculation so that this condition is not
sufficiently stringent.

We could proceed by constructing the approximate paths associated to
each of the unperturbed paths of Figure 2. Instead, we will
use the analysis of section 5, above, to obtain important
qualitative information about the solutions.
For a given background geodesic,
that analysis instructs us, first, to consider an initial value
type solution, corresponding to a photon
emitted backwards in time from the observer
in the direction of the background geodesic.
We know from Pyne and Birkinshaw (1993)
that the initial value calculation returns a deflection
at the lens equal to $L\alpha$, where $\alpha$ is the usual
lens deflection angle appropriate to the background path used,
$\alpha=4m/L\sin\theta$. The method then would correct for
the distance between $x^{(0)i}(0)$ and the origin, essentially by
(vectorially) subtracting this distance from the deflection
computed at the lens plane. The result, considered as the net
linear deflection undergone by the photon, would then
determine the position of the emitter on the observer's sky by
applying to it the inverse Jacobi operator, basically $1/2L$. The
paths constructed in this manner are shown in Figure 3.

Comparison of Figures 2 and 3, however, reveals an important
distinction between the ``correct" solution,
$x^{\mu}_B$, and the two extreme
solutions, $x^{\mu}_A$ and $x^{\mu}_C$.
The two extreme solutions sample a lens potential
totally unlike that felt by the background paths they were
constructed from. In contrast, the solution marked $x^{\mu}_B$
feels essentially the same lens potential as that felt by
the background geodesic it was constructed from, $x^{(0)}_B$.
We could, for instance, conjecture that the two paths we seek
are those generated by the two background paths which
minimize

$$\left| \ \int\limits_{x^{\mu}}|\phi | \, d\lambda -\int
\limits_{x^{(0)\mu}} |\phi |\,
d\lambda \ \right| \eqno{(58)}$$

\noindent
where $x^{\mu}$ is the perturbed path associated to $x^{(0)\mu}$.
Of course, (58) is mostly heuristic. We are not
proposing this expression as a general error functional, assigning some
significant real number measure of the error involved in using
any particular background geodesic. \footnote{$^*$}{
Nevertheless, such a functional
should not be too difficult to find. The error in the
perturbative geodesic expansion is a result of error in the
approximate solution to Einstein's equation itself, which we do not
consider, error from linearization of the Christoffel
sysmbols and error from the truncation of the Taylor expansions used to
express these quantities along the background geodesic (Pyne and
Birkinshaw 1993). Simple matrix methods and the well known
remainder term for Taylor's theorem can furnish crude bounds
on these last two sources of error.} It does happen to
suffice here, however.

It is important to recognize that the above
is exactly the sort of reasoning which must be
applied in making rough consistency judgements in an initial value
calculation. In fact, there is an analogous background path
freedom in those calculations as well: by changing the initial
data, an infinite number of background paths can be made to
generate an infinite number of distinct perturbed geodesics
all of which pass though a given point with the same tangent vector.
Of course, there is a unique actual geodesic which passes through
this point with the given tangent. It is nearly always assumed
in perturbative calculations that the specific perturbed geodesic
under investigation is, if not the best, at least an adequate
approximation to the actual geodesic. In fact, we often have no
more reason to expect this in the common initial value cases as we
do for the fixed endpoint scenarios. It is a common hope, however,
that forewarned is largely forearmed.

Suppose, now, that we have applied some error minimization
and gained the usual lens solution.
Would we have gained the usual lens time-delay formula?
The answer is no, because the usual time-delay formula (see, e.g.
Schneider, Ehlers, and Falco 1993, Chapter 4)
contains a second order term,
the geometric delay term, which is proportional to the square of the
lens angle, itself a first order quantity. In fact,
as we have already noted, the lens
perturbation we are considering, when used in (10), immediately
produces the Shapiro delay, but contributes nothing else.
{}From a perturbation-theoretic
point of view, the inclusion of the geometric delay term in the
standard treatments is quite {\it ad hoc}. In Seljak (1994),
for instance, the spatial components of the lensed photon path
are solved for to linear order in the perturbing potential. This
projected path is then ``lifted" into the time domain by imposing
$ds^2=0$ at second order. We do not intend this remark as a critique
on the usual treatments: it is possible to perform
the entire calculation at second order and show that only the
usual terms are important. We bring this issue up
in order to emphasize that there are astrophysical
instances when {\it a priori} second order terms contribute numerically
as importantly as first order terms. A nice example of this is the
recent work of Frieman, Harari, and Surpi (1995).

\vskip0.5truein
\noindent
{\bf 7. Summary}

\nobreak
We have presented an equation (29) for the time delay measured by
two antennae observing a distant source through an adiabatic
background of gravitational waves on Einstein-de Sitter spacetime.
We have used the equation to determine the pattern of source motions
on the sky induced by a background of gravitational radiation.
We have also shown how these results may be extended to the curved FRW
spacetimes. Our results are immediately applicable to situations,
such as that considered here, for which the standard formulae
of Kristian and Sachs (1965) do not apply, and so represent
an important contribution to the theory of observations in
perturbed spacetimes.

\vskip0.5truein
\noindent
{\bf Acknowledgements}

We thank R. M. Campbell and Sean Carroll for helpful discussions.
This work was supported by the National Science Foundation under grants
AST90-05038 and AST-9217784.

\vfill\eject

\noindent
{\bf References}

\vskip 12pt
\normalbaselineskip=8pt plus0pt minus0pt
                            \parskip 0pt

\def\ref#1  {\noindent \hangindent=24.0pt \hangafter=1 {#1} \par}
\def\vol#1  {{\bf {#1}{\rm,}\ }}
\ref{Abbott, L., \& Wise, M. 1984, Nucl. Phys., B244, 541}
\ref{Carr, B. J. 1980, A\&A, 89, 6}
\ref{Einstein, A. 1916, {\it Preuss. Akad. Wiss. Berlin Sitzber.},
688}
\ref{Eubanks, T. M., \& Matsakis, D. N. 1994, unpublished}
\ref{Fabbri, R., \& Pollock, M. 1983, Phys. Lett., 125B, 445}
\ref{Frieman, J. A., Harari, D. D., \& Surpi, G. C. 1995,
preprint/a-ph9405015}
\ref{Gradshteyn, I. S., \& Ryzhik, I. M. 1994, Table of Integrals,
Series, and Products (San Diego: Academic Press)}
\ref{Gwinn, C. G., Pyne, T., Birkinshaw, M., Eubanks, T. M., \&
Matsakis, D. N. 1995, in preparation}
\ref{Hawking, S. W. 1966, ApJ, 145, 544}
\ref{Kristian, J., \& Sachs, R. K. 1965, ApJ, 143, 379}
\ref{Linder, E. V. 1988a, ApJ, 326, 517}
\ref{Linder, E. V. 1988b, ApJ, 328, 77}
\ref{Mukhanov, V. F., Feldman, H. A., \& Brandenberger, R. H. 1992,
Phys. Rep., 215 Nos. 5 \& 6, 203}
\ref{Pyne, T. \& Birkinshaw, M. 1993, ApJ, 415, 459}
\ref{Pyne, T. \& Birkinshaw, M. 1995, ApJ, submitted}
\ref{Romani, R. W., \& Taylor, J. H. 1983, ApJ, 265, L35}
\ref{Rubakov, V. A., Sazhin, M., \& Veryaskin, A. 1982,
Phys. Lett., 115B, 189}
\ref{Sachs, R. K., \& Wolfe, A. M. 1967, ApJ, 147, 73}
\ref{Schneider, P., Ehlers, J., \& Falco, E.~E. 1993,
Gravitational Lensing (Berlin: Springer-Verlag)}
\ref{Seljak, U. 1994, ApJ, 436, 509}
\ref{Taylor, J. H. 1992, in Proceedings of the 13th International
Conference on General Relativity and Gravitation, eds.
R. J. Gleiser, C. N. Kozameh, \& O. M. Moreschi (Philadelphia:
Institute of Physics Publishing)}
\ref{Taylor, J. H. 1987, in Proceedings of the 13th Texas
Symposium on Relativistic Astrophysics, ed. M. P. Ulmer
(Singapore: World Scientific)}
\ref{Taylor, J. H., \& Weisberg, J. M. 1989, ApJ, 345, 434}
\ref{Thompson, A. R., Moran, J. M., \& Swenson, G. W. Jr. 1986,
Interferometry and Synthesis in Radio Astronomy (New York: John
Wiley and Sons)}
\ref{Thorne, K. S. 1987, in Three Hundred Years of Gravitation,
eds. S. W. Hawking \& W. Israel (Cambridge: Cambridge University
Press)}
\ref{Weinberg, S. 1972, Gravitation and Cosmology (New
York: John Wiley and Sons)}

\normalbaselineskip=24pt plus0pt minus0pt
                  \parskip 12.0pt
\vfill\eject

\noindent
{\bf Figure 1.} The geometry of Section 3. In the perturbed spacetime,
light from a source at $q$ travels along the null geodesic $x^{\mu}_A$
to an antenna at $w^{\prime}_A$. The antenna is separated from the
origin of spatial co-ordinates by the spatial vector ${\vec r}_A$.
$\tau_A$ is the line of time translates of $w^{\prime}_A$.
The null geodesic of the background $x^{(0)\mu}_A$ joins the source
to a point $w_A\in \tau_A$.

\noindent
{\bf Figure 2.} The background paths of Section 6, projected into the
$xy$-plane. $R_s$ is the Schwarzschild radius of the lensing mass.
The arrows point in the direction of photon travel. $\theta_E$ is
the Einstein angle of the lens.

\noindent
{\bf Figure 3.} The solution paths for the fixed-endpoint problem along
the background paths of Figure 2. $R_s$ is the Schwarzschild radius of the
lensing mass.
The arrows point in the direction of photon travel. $\theta_E$ is
the Einstein angle of the lens.

\vfill\eject

\end